\documentclass[aps,preprint,groupedaddress,showpacs]{revtex4}
\begin{document}

\title{Optical Properties of the Spin-Ladder Compound
 Sr$_{14}$Cu$_{24}$O$_{41}$}
\author{Z. V. Popovi\'c $^{a}$, M. J. Konstantinovi\'c $^{b}$,
  V. A. Ivanov $^{a}$, O. P. Khuong $^{a}$, R. Gaji\'c $^{c}$,
  A. Vietkin $^{d}$ and V. V. Moshchalkov $^{a}$}

\affiliation{ $^a$ Laboratorium voor Vaste-Stoffysica en
Magnetisme, K. U. Leuven, Celestijnenlaan 200D, B-3001 Leuven,
Belgium } \affiliation{$^b$ Max-Planck-Institut f\"ur
    Festk\"orperforschung, Heisenbergstrasse 1,D-70569 Stuttgart, Germany}
\affiliation{$^c$ Institute of Physics, P.O.Box 68, 11080
Belgrade, Yugoslavia} \affiliation{$^d$ Physics Department, Moscow
State University, 119899 Moscow, Russia}

\begin{abstract}
We report the measurements of the pseudodielectric function,
 far-infrared reflectivity and Raman scattering spectra in
  Sr$_{14}$Cu$_{24}$O$_{41}$ single crystal. We
study the lattice and the spin dynamics of the Cu$_2$O$_3$ spin
ladders and CuO$_2$ chains of this compound. The ellipsometric and
the optical reflectivity measurements yield the gap values of 1.4
eV, 1.86 eV, 2.34 eV (2.5 eV) for the ladders (chains) along the
{\bf c}-axis
 and 2.4 eV along the {\bf a}-axis.
 The electronic structure of the Cu$_2$O$_3$ ladders is analyzed
 using tight-binding approach for the correlated electron systems.
 The correlation gap value of 1.4 eV is calculated with the transfer energy
 (hopping) parameters $t=t_{0}$=0.26 eV, along and perpendicular to legs,
 $t_{xy}$=0.026 eV (interladder hopping) and U=2.1
eV, as a Coulomb repulsion.
  The optical parameters of the infrared active
phonons and plasmons are obtained by oscillator fitting procedure
of the reflectivity spectra. Raman scattering spectra are measured
at different temperatures using different laser line energies. The
two-magnon peak is observed at about 2880 cm$^{-1}$. At
temperatures below 150 K the new infrared and Raman modes appear
 due to the charge ordering.
\end{abstract}
 \maketitle

\section{Introduction}
The Sr$_{14}$Cu$_{24}$O$_{41}$ compound is one of the three stable
phases in the Sr-Cu-O system which can be synthesized under
ambient pressure. The other two stable phases are Sr$_2$CuO$_3$,
which has simple chains of Cu ions, and SrCuO$_2$ with zigzag
chains of Cu ions.  This oxide has unique crystal structure based
on two sublattices; one of them consists of Cu$_2$O$_3$ two-leg
ladders and the second one is formed by CuO$_2$ chains. These two
sublattices
 are incompatible along one crystallographic direction, thus resulting in
an 1D incommensurate structure \cite {a1}.  According to the
structural analysis \cite {a1}, the ladder sublattice is
face-centered-orthorhombic (space group $Fmmm$) with a lattice
parameters a=1.1459 nm, b=1.3368 nm and c$_{Ladder}$=0.3931
 nm. There are two ladder layers with two ladders per unit cell, Fig. 1. The
chain sublattice is A-centered orthorhombic (space group $Amma$),
with nearly the same {\bf a} and {\bf b} axes but different {\bf
c} axis, c$_{Chain}$=0.2749 nm. However,
Sr$_{14}$Cu$_{24}$O$_{41}$ can be considered as nearly
commensurate structure at 7xc$_{Ladder}$=2.7372 nm and
10xc$_{Chain}$=2.7534 nm.  The schematic illustration of
Sr$_{14}$Cu$_{24}$O$_{41}$ crystal structure is given in Fig. 1.

The physical properties of Sr$_{14}$Cu$_{24}$O$_{41}$ have
attracted a lot of attention recently
 \cite {a2,a3,a4,a5,a6,a7,a8,a9,a10,a11,a12}, in connection with
the rich physics associated with the S=1/2 Heisenberg
antiferromagnetic quasi one-dimensional (1D) structures and the
discovery of superconductivity in
Sr$_{0.4}$Ca$_{13.6}$Cu$_{24}$O$_{41}$ under high pressure \cite
{a2}. The progress in this field has been summarized in Ref.
\cite{a13}.
 Various magnetic \cite {a3,a4,a5}, NMR \cite {a6,a7} and
neutron scattering \cite {a8,a9} measurements, showed that
Sr$_{14}$Cu$_{24}$O$_{41}$ has two kinds of magnetic gaps. The
first one, $\Delta_L$ = 32.5 meV, is attributed to the
singlet-triplet excitation in the Cu$_2$O$_3$ spin-ladders, with
the exchange energies
 along the legs
(J=130 meV) and the rungs (J${_0}$=72 meV) \cite {a8}.
  The second one, $\Delta_C$ = 11.5 meV, is argued to arise
  from the spin dimer formation in the CuO$_2$ chains \cite {a8}, with an
antiferromagnetic intradimer coupling  J${_1}$=11.2 meV.
 Similar
ratio of the superexchange interaction energies,
 J${_0}$/J $\sim$ 0.5, as well as
the magnitude of J${_0}$=(950$ \pm$300) K, is found in the
$^{17}$O and $^{63}$Cu NMR measurements \cite {a6}.

The Raman spectra of Sr$_{14}$Cu$_{24}$O$_{41}$ were measured previously
 \cite {a14,a15}. From the comparison between Raman spectra of the
various layered tetragonal cuprates, Abrashev {\it et al.}. \cite
{a14}
 concluded that the main contribution to the spectra comes from the Raman
forbidden-infrared active LO phonons and the two-magnon
scattering. Furthermore, Sugai {\it et al.} \cite {a15} argued
that, besides strong two-magnon features, some low-frequency modes
in the Raman spectra are also magnetic in origin, since they have
similar energies to those found in the neutron scattering
experiments \cite {a8}. Still, for the proper identification of
the magnetic modes, the temperature dependent Raman spectra, as
well as the spectra in magnetic field, are required. Also,
detailed analysis of the lattice dynamics and comparison between
Raman and infrared (IR) spectra are indispensable due to the
incommensurability of the structure. Therefore, we present here
the Raman and IR spectra at various
 temperatures between 5 and 300 K in order
 to make more complete assignment of the vibrational modes in
  Sr$_{14}$Cu$_{24}$O$_{41}$.  The Raman spectra are also measured
 under resonant conditions, with a laser light energy close to gap values. The correlated electron
tight-binding model of electronic structure is used to estimate
the hopping parameters, in fact adjusted to the measured gaps and exchange energies.

\section{Experimental details}
The present work was performed on (010) oriented single crystal
plates with dimensions typically about 5 x 1 x 6 mm$^3$ in the
{\bf a}, {\bf b} and {\bf c} axes, respectively.  The infrared
measurements were carried out with a BOMEM DA-8 FIR spectrometer.
A DTGS pyroelectric detector was used to cover the wave number
region from 100 to 700 cm$^{-1}$; a liquid nitrogen cooled HgCdTe
detector was used from 500 to 1500 cm$^{-1}$. The spectra were
collected with the 2 cm$^{-1}$ resolution.  The low temperature
reflectivity spectra in the range from 30 to 5000 cm$^{-1}$ were
measured using Bruker IFS 133v FIR-spectrometer with
Oxford-Cryostat.  The Raman spectra were recorded in the
backscattering configuration using micro- and macro- Raman systems
with Dilor triple monochromator including liquid nitrogen cooled
CCD-detector. An Ar- and Kr -ion lasers were used as an excitation
source. We measured the pseudodielectric function with a help of a
rotating-analyzer ellipsometer. We used a Xe-lamp as a light
source, a double monochromator with 1200 lines/mm gratings and an
S20 photomultiplier tube as a detector. The polarizer and analyzer
were Rochon prisms.  The measurements were performed in the
1.6-5.6 eV energy range. Optical reflectivity spectra were
measured at room temperature in the 200-2000 nm spectral range
using Perkin-Elmer Lambda 19 spectrophotometer.

\section{Electronic structure}

The electronic structure of Sr$_{14}$Cu$_{24}$O$_{41}$ is
calculated using the tight-binding method for correlated electrons
\cite {a16}. Recent exact diagonalization and a variational Monte
Carlo simulations revealed that electronic structure of
 Sr$_{14}$Cu$_{24}$O$_{41}$ is well described by single ladder energy
spectrum \cite {a17}. It means that electron energy dispersions
are governed mainly by electrons in the ladder. According to the
ARPES measurements $\left[ 10\right] $, the chains contribute to
the electronic structure of Sr$_{14}$Cu$_{24}$O$_{41}$ with a
dispersionless band. Without entering any lengthy discussions
about the substance nonstoichiometry and the carrier transport
between chains and ladders, we assume further on that the ladder
unit Cu$_2^{1+n}$O$_3^{2-}$ has total charge $-2$. In other words,
there is one hole, $n=1$, per copper ion in the ladder for
negligibly small hybridization of its $d_{x^2-y^2}-$orbitals
with the $p_y-$%
orbital of intermediate oxygens. The angle between Cu atoms of
neighboring ladders is near right angle $\left( 88.7^{\circ
}\right) $, Fig. 1. In our consideration of the electronic
structure we assumed that the directions between the nearest
neighbor Cu ions form an ideal right angle.

The Hamiltonian for the correlated copper holes in the ladder with
two rungs, $a-b$ and $c-d$, per a unit cell can be written as

\begin{eqnarray}
H &=&-2t\sum\limits_{p,\sigma }\cos p_y\left[ a_\sigma ^{+}\left(
p\right) a_\sigma ^{}\left( p\right) +b_\sigma ^{+}\left( p\right)
b_\sigma ^{}\left( p\right) +c_\sigma ^{+}\left( p\right) c_\sigma
^{}\left( p\right) +d_\sigma ^{+}\left( p\right) d_\sigma
^{}\left( p\right) \right] \nonumber\\ &&-t_0\sum\limits_{p,\sigma
}\left[ a_\sigma ^{+}\left( p\right) b_\sigma ^{}\left( p\right)
+c_\sigma ^{+}\left( p\right) d_\sigma ^{}\left( p\right)
+H.c.\right] \nonumber \\
&&-t_{xy}\sum\limits_{p,\sigma }\left[ e^{-i\sqrt{2}p_x}+e^{-i(\sqrt{2}%
p_x+p_y)}\right] \left[ a_\sigma ^{+}\left( p\right) d_\sigma
^{}\left( p\right) +H.c.\right]   \nonumber \\
&&-t_{xy}\sum\limits_{p,\sigma }\left( 1+e^{-ip_y}\right) \left[
b_\sigma ^{+}\left( p\right) c_\sigma ^{}\left( p\right)
+H.c.\right] +U\sum\limits_{i=a,b,c,d}n_{\uparrow
i}^jn_{\downarrow i}^j-\mu \sum\limits_{i=a,b,c,d}n_i^j, \nonumber
\\ &&
\end{eqnarray}

where $a,b,c,d$ represent chains, (see Fig.1.(a)), $t$ $(t_0)$
are the values of the carrier hopping along legs (rungs), $t_{xy\text{ }}$%
is a hopping amplitude between ladders, $U$ is the
Anderson-Hubbard repulsion, and $\mu$ is the chemical potential
and other notations as usual. The $x$ and $y$ axis of the
reference system are taken along the {\bf a} and the {\bf c}
crystallographic directions, respectively. Applying the
$X-$operator machinery \cite {a18}, the correlation split energy
bands are governed by zeros of the inverse Green's function in the
first perturbation order with respect to tunneling matrix:
\begin{equation}
\widehat{D}_p^{-1}\left( \omega \right) =\left(
\begin{tabular}{ll}
A$_{a-b}$ & B \\
B$^{*}$ & A$_{c-d}$%
\end{tabular}
\right) ,
\end{equation}
where
\begin{eqnarray*}
\widehat{A}_{a-b} &=&
\begin{tabular}{l}
$a$ $\{
\begin{tabular}{l}
0+ \\ -2
\end{tabular}
$ \\ $b$ $\{
\begin{tabular}{l}
0+ \\ -2
\end{tabular}
$
\end{tabular}
\left(
\begin{array}{llll}
\frac{-i\omega _n-\mu }{f_{0+}}+r & r & -t_0 & -t_0 \\ r &
\frac{-i\omega _n-\mu +U}{f_{-2}}+r & -t_0 & -t_0 \\ -t_0 & -t_0 &
\frac{-i\omega _n-\mu }{f_{0+}}+r & r \\ -t_0 & -t_0 & r &
\frac{-i\omega _n-\mu +U}{f_{-2}}+r
\end{array}
\right) , \\ \widehat{B} &=&\left(
\begin{array}{llll}
0 & 0 & D & D \\ 0 & 0 & D & D \\ C & C & 0 & 0 \\ C & C & 0 & 0
\end{array}
\right) \text{ } \\ \text{with }r &=&-2t\cos p_y,C=-t_{xy}\left(
1+e^{-ip_y}\right) ,D=-t_{xy}\left[
e^{-i\sqrt{2}p_x}+e^{-i(\sqrt{2}p_x+p_y)}\right] \text{.}
\end{eqnarray*}

Here the correlation factors $f_{0+}$, $f_{-2}$ are determined by
fermion occupation $n$ per
site. Namely, for the considered paramagnetic phase they are $f_{0+}=1-n/2,$ $%
f_{-2}=n/2$ and all equal to $1/2$ ($n=1$). After an analytical
continuation, $i\omega _n\rightarrow \xi +i\delta $, in Eq.
$\left| \widehat{D}_p^{-1}\left( \omega \right) \right| =0$ one
can find the bonding /antibonding correlation energy dispersions
as follows:

\begin{equation}
\xi _B^{\pm }\left( p\right) =\frac 12\left[ \varepsilon _p^{1,2}+\sqrt{%
\left( \varepsilon _p^{1,2}\right) ^2+U^2}\right] ,
\end{equation}\\
\begin{equation}
\xi _A^{\pm }\left( p\right)  =\frac 12\left[ \varepsilon _p^{1,2}-\sqrt{%
\left( \varepsilon _p^{1,2}\right) ^2+U^2}\right] ,
\end{equation}\\
\text{where }$\varepsilon _p^{1,2} =-2t\cos p_y\pm
\sqrt{t_0^2+\left(
2t_{xy}\cos \frac{p_y}2\right) ^2\pm 4t_0t_{xy}\cos \frac{p_y}2\cos \frac{%
\sqrt{2}p_x}2}$.  \nonumber

For derivation of these energy dispersions from the eight-by-eight
fold secular equation (see Eq.(2)) it was useful to apply the
theorem about the decomposition of determinant with respect to
diagonal elements (see Appendix A in Ref. \cite {a19}).
 The subbands $\xi _B^{-}$ and $\xi
_A^{-}$ are completely occupied by carriers with concentration
$n=1$ per copper site of ladder for the chosen chemical potential
$\mu =U/2$. The nearest unoccupied energy band is $\xi _B^{+}$ and
the correlation gap in electronic structure can be estimated as
\begin{eqnarray}
\Delta_{corr}  &=&\min \xi _B^{+}\left( p\right) -\max \xi
_A^{-}\left( p\right) =
 \\
&&\frac 12\left[ \sqrt{\left( t_0+2t-2t_{xy}\right)
^2+U^2}+\sqrt{\left( t_0+2t\right) ^2+U^2}\right] -\left(
t_0+2t+t_{xy}\right) .  \nonumber
\end{eqnarray}
For the dimensionless energies, $\omega _p^{+,-}=\varepsilon _p^{1,2}/2t$, $%
\tau _0=t_0/2t, \tau =t_{xy}/2t,$ the non-correlated electron
density of
states per spin, $\rho $($\varepsilon $)$=\sum\limits_{p_{x,}p_y}$[$\delta $%
($\varepsilon -\omega _p^{+}$)$+\delta $($\varepsilon -\omega
_p^{-}$)], for the unit cell volume, is defined analytically as
follows

\begin{eqnarray}
\rho _0^\alpha \left( -1-2\tau +\alpha \tau _0\leq \varepsilon
^{}\leq -1+2\tau -\tau _0\right)  &=&\frac 4{\pi ^2\sqrt{k_\alpha
\tau }}K\left( q_\alpha \right) ,  \nonumber \\ \rho _0^\alpha
\left( -1+2\tau +\alpha \tau _0\leq \varepsilon ^{}\leq \frac
12+\tau +\alpha \tau _0\right)  &=&\frac 4{\pi ^2q_\alpha
\sqrt{k_\alpha \tau }}F\left( \arcsin a_\alpha ;\frac 1{q_\alpha
}\right) ,   \\ \rho _0^\alpha \left( -1+2\tau +\alpha \tau _0\leq
\varepsilon ^{}\leq
1+\alpha \tau _0\right)  &=&\frac 4{\pi ^2q_\alpha \sqrt{k_\alpha \tau }%
}K\left( \frac 1{q_\alpha }\right) .  \nonumber
\end{eqnarray}
Eqs.(6) represent the analytical expressions for the electron
density of states via elliptic integrals $F$ and $K$ of the 1-st
kind in the Legendre normal form with modulus $q_\alpha
=\sqrt{\left[ 2\tau \left( \tau +k_\alpha \right) +1-\left(
\varepsilon +\alpha \tau _0\right) ^2\right] /k_\alpha \tau }/2$,
argument $a_\alpha =\sqrt{\left[ 2\tau \left( \tau +k_\alpha
\right) +1-\left( \varepsilon +\alpha \tau _0\right) ^2\right]
/\left[ \left( 1+\varepsilon +\alpha \tau _0\right) \left( \tau
+k_\alpha \right) k_\alpha \right] }$, where $k_\alpha =\sqrt{\tau
^2+2\left( 1-\varepsilon -\alpha \tau _0\right) }$ and $\alpha
=\pm $.

The electron-electron repulsion, $U$, splits the density of the
non-correlated electronic states $\rho _0$. The correlated
electron density of states is
\begin{eqnarray}
\rho \left( \varepsilon \right)  &=&\frac 2{1+\frac{\xi _{+}^{^{\prime }}}{%
\sqrt{\left( \xi _{+}^{^{\prime }}\right) ^2+\left( \frac U{2t}\right) ^2}}%
}\rho _0\left( \xi _{+}^{^{\prime }}\right) +\frac 2{1-\frac{\xi
_{-}^{^{\prime }}}{\sqrt{\left( \xi _{-}^{^{\prime }}\right)
^2+\left( \frac U{2t}\right) ^2}}}\rho _0\left( \xi _{-}^{^{\prime
}}\right) ,\text{where }  \\ \text{ }\xi _{\pm }^{^{\prime }}
&=&\frac{\xi _{\pm }^{^2}-\left( \frac U{2t}\right) ^2}{\xi _{\pm
}^{}}+S  \nonumber \\ \text{and }S &=&\frac 14\left[ \sqrt{\left(
1+2\tau +\tau _0\right) ^2+\left( \frac Ut\right) ^2}-\sqrt{\left(
1+\tau _0\right) ^2+\left( \frac Ut\right) ^2}-2\tau \right] ,
\nonumber
\end{eqnarray}
are expressed via dimensionless correlated energies $\xi _{\pm
}^{^{}}\equiv \xi ^{\pm }\left( p\right) /2t$. With the help of
Eq.(7) and Eqs.(6) one can calculate the correlated electron
density of states for corresponding energy ranges. Its explicit
form naturally includes the first kind elliptic integrals. The
results of the calculations are plotted in Fig. 2.

The overlap of the energy ranges for the electronic dispersions,
Eq.(3),
leads to the special features of the correlated electronic structure at $%
L_5=1/2\left( -1-2\tau +\tau _0\right) -S-1/2\sqrt{\left( 1+2\tau
+\tau _0\right) ^2+\left( U/t\right) ^2}\ $and $L_7=1/2\left(
1/2+\tau +\tau
_0\right) -S-1/2\sqrt{\left( 1/2+\tau +\tau _0\right) ^2+\left( U/t\right) ^2%
}$ in the lower correlated band. Logarithmic divergencies
inside the band  at $L_2=1/2\left( -1+2\tau -\tau _0\right) -S-1/2\sqrt{%
\left( -1+2\tau -\tau _0\right) ^2+\left( U/t\right) ^2}$,
$L_4=1/2\left(
1-\tau _0\right) -S-1/2\sqrt{\left( 1-\tau _0\right) ^2+\left( U/t\right) ^2}%
$ , $L_6=1/2\left( -1+2\tau +\tau _0\right) -S-1/2\sqrt{\left(
-1+2\tau
+\tau _0\right) ^2+\left( U/t\right) ^2}$ and at the correlated band edge $%
L_8=1/2\left( 1+\tau _0\right) -S-1/2\sqrt{\left( 1+\tau _0\right)
^2+\left( U/t\right) ^2}$ are clear manifestations of the 2D
electronic structure of Sr$_{14}$Cu$_{24}$O$_{41}$ compound. We
would like to emphasize that in the one-dimensional case $\left(
t_{xy}\rightarrow 0\right)$ the electron density of states is
taking features of a single spin-ladder without any logarithmic
peaks, then the divergencies become square-root like and they are
located at the band edges, $\varepsilon =\pm 1$: $\rho \left(
\varepsilon \right) =8K\left( 0\right) /\pi ^2\sqrt{1-\varepsilon
^2}=4/\pi \sqrt{1-\varepsilon ^2}$.

\section{Experimental results}
The dielectric function $\epsilon_2$ of Sr$_{14}$Cu$_{24}$O$_{41}$
is shown in Fig.3 in the spectral region from 1.6 eV to 5.5 eV.
 These spectra were computed from the measured
Fourier coefficients using the equations for an isotropic case.
Consequently, $\epsilon_2$ represents a complicated average of the
projections of the dielectric tensor on the sample surface.  We
presented the spectra of the (010) surface taken with the {\bf
a}-axis, Fig. 3(a) and {\bf c} axis, Fig. 3(b), in the plane of
incidence (PI).  According to the prescription given by Aspnes
\cite {a20}, we attribute these components to the components of
the dielectric tensor $\epsilon_2^{aa}$ and $\epsilon_2^{cc}$. The
bands with the energies of 2.4, 4.1, and 4.7 eV for the {\bf
a}-axis and at about 1.86, 2.34, 2.5, and 4.3 eV are found for the
{\bf c} axis in the plane of incidence, respectively.

Inset (a) in Fig. 3 shows reflectivity spectra of
Sr$_{14}$Cu$_{24}$O$_{41}$. These spectra are calculated from
measured dielectric functions $\epsilon_1$ and $\epsilon_2$. Inset
(b) in Fig. 3 represents the unpolarized optical reflectivity of
Sr$_{14}$Cu$_{24}$O$_{41}$ measured at room temperature. In
addition to peaks, previously observed in ellipsometric
measurements, a new peak at about 1.4 eV appears.

    The room temperature polarized far-infrared
reflectivity spectra of Sr$_{14}$Cu$_{24}$O$_{41}$ are given in
Fig. 4.  The open circles are the experimental data and the solid
lines represent the spectra computed using a four-parameter model
for the dielectric constant:

\begin{equation}
\epsilon(\omega)=\epsilon_{\infty} (\prod_{j=1}^{n}
\frac{\omega_{LO,j}^2-\omega^2+\imath
\gamma_{LO,j}\omega}{\omega_{TO,j}^2-\omega^2+\imath\gamma_{TO,j}\omega}-\frac{\omega_p^2}
{\omega(\omega-\imath\tau^{-1})}), \label{7}
\end{equation}

where $\omega_{LO,j}$ and $\omega_{TO,j}$ are longitudinal and transverse frequencies
of the j$^{th}$
oscillator, $\gamma_{LO,j}$ and $\gamma_{TO,j}$ are their corresponding
dampings, $\omega_p$ is the plasma frequency, $\tau$ is the free-carrier relaxation time and
$\epsilon_{\infty}$ is the
high-frequency dielectric constant.

The best fit parameters are given in Table I.  The agreement
between observed and calculated reflectivity spectra is rather
good.  For the {\bf E}$||${\bf a} polarization, eight oscillators
with TO frequencies at about 164, 194, 219, 249, 283.5, 310.4, 554
and 623 cm$^{-1}$ are clearly seen.  In the {\bf E}$||${\bf c}
polarization, Fig. 4(b), nine oscillators at 135, 148, 253, 293,
345, 486, 540, 596 and 620 cm$^{-1}$ are observed.  Besides
phonons, our model includes the Drude expression for light
scattering on free carriers.  We obtained the plasma frequency at
about 4000 cm$^{-1}$ (1000 cm$^{-1}$) for the {\bf E}$||${\bf c}
({\bf E}$||${\bf a}) polarizations.

  The room temperature Raman spectra of
Sr$_{14}$Cu$_{24}$O$_{41}$, for (aa) and (cc) polarized
configurations are presented
 in Figs. 5(a) and 5(b).  These spectra  consist of only
A$_g$ symmetry modes.  Four modes at 246, 302, 548, and 582
cm$^{-1}$ are clearly seen.  The low temperature (cc) Raman
spectra are given in Figs. 5(c)-(g).  By lowering temperature
below 200 K, the modes narrow and in addition the new modes
appear. We will discuss them later on.
  The Raman spectra of Sr$_{14}$Cu$_{24}$O$_{41}$, excited by
 different lines of Ar and Kr lasers at 8 K, are shown in Fig. 6 for the (cc)
and (aa) polarized configurations in the spectral ranges: (a) from 700 to
1400 cm$^{-1}$, (b) from 1675 to 1975 cm$^{-1}$ and (c) from 2600 to
3300 cm$^{-1}$.  Anticipating our conclusions, we divide the
Raman spectra in three different energy regions:  one phonon (0-700
cm$^{-1}$), two-phonon (700-1400 cm$^{-1}$) and two-magnon region
(above 1500 cm$^{-1}$).  The mode at about 2900 cm$^{-1}$ and a
broad structure at about 1900 cm$^{-1}$ are magnetic in origin,
according to their intensity and frequency
dependence as a function of the temperature, see Fig. 7.

\section{Discussion}
The average unit cell of Sr$_{14}$Cu$_{24}$O$_{41}$ consists of four formula
units with 316 atoms in all. Since there is a large
 number of atoms in the unit cell, we can expect a very large
number of optically active modes. Consequently, the lattice
dynamical calculation is practically impossible. All atoms have
4(e) position symmetry of
 $Pcc2$ $(C_{2v}^3)$ space group \cite {a1}. Factor-group-analysis (FGA) yields the following
 distribution of vibrational modes:
\begin{equation}
\Gamma_{Sr_{14}Cu_{24}O_{41}}= 237A_1 ({\bf E}||{\bf c},
 aa, bb, cc) + 237A_2 (ab) + 237B_1 ({\bf E}||{\bf a}, ac)
+ 237B_2 ({\bf E}||{\bf b}, bc)
\label{8}
\end{equation}
According to this representation one can expect 948 modes which
are both Raman and infrared active. Experimentally, the number of
observed modes is less then ten for each polarization. Because of
that, we consider separately the contribution of each sublattice
unit.
  As mentioned earlier, the space group of ladder
sublattice is $Fmmn$ $(D_{2h}^{23})$.  The site symmetries of Sr,
Cu,
 O$_1$ and O$_2$ atoms
are (8h), (8g), (8g) and (4b), respectively.  The FGA for the
ladder structure (Sr$_2$Cu$_2$O$_3$) yields \cite {a21}:

$\Gamma_{Ladder}$ = 3$A_g + 3B_{1g} + 2B_{2g} + B_{3g} + 4B_{1u} +
4B_{2u} + 4B_{3u}$

The space group of a chain sublattice is $Amma$ $(D_{2h}^{17})$.
The site symmetries of Cu and O atoms are (4c) and (8f).  The FGA
gives for the chain structure:

$\Gamma_{Chain}$ = $3A_g + 3B_{1g} + 2B_{2g} + B_{3g} + A_u +
2B_{1u} + 3B_{2u} +
 3B_{3u}$,

Since $Amma$ is not a standard setting for $D_{2h}^{17}$ space
group ($Cmcm$) we use $C_s^{xy}$ site symmetry for oxygen atoms
and $C_{2v}^x$ symmetry for Cu atoms in above representations.
Thus, the total number of vibrational modes from both sub-units
is:
\begin{equation}
\Gamma =6A_g(aa,
bb,cc)+6B_{1g}(ab)+4B_{2g}(ac)+2B_{3g}(bc)+A_u+6B_{1u}({\bf
E}||{\bf c})+7B_{2u}({\bf E}||{\bf b})+7B_{3u}({\bf E}||{\bf a})
\label{9}
\end{equation}
According to this analysis we should expect 6A$_g$ modes; one mode
from vibrations of the Sr atoms, two modes which originate from
vibrations of Cu atoms and other three A$_g$ modes are due to
oxygen vibrations. In order to assign the observed A$_g$ modes we
compare our spectra with the corresponding spectra of the Cu-O
based materials with similar structural units as in
Sr$_{14}$Cu$_{24}$O$_{41}$. For example, in SrCuO$_2$ \cite {a22}
and YBa$_2$Cu$_4$O$_8$ \cite {a23} the Cu-O double layers exist
and resemble the one leg of the ladder structure in
Sr$_{14}$Cu$_{24}$O$_{41}$. The Cu-O chains, formed from copper
oxide squares with the common edges,
 as in Sr$_{14}$Cu$_{24}$O$_{41}$, are also present in CuO
\cite {a24} and CuGeO$_3$ \cite {a25}. Thus, the lowest frequency
mode in Fig. 5(a) at 246 cm$^{-1}$ can be assigned to vibrations
of the Cu ladder atoms (see Fig. 1).  The corresponding A$_g$
 mode of copper
vibrations in YBa$_2$Cu$_4$O$_8$ (SrCuO$_2$) appears at 250 (263)
cm$^{-1}$. The next mode is found at 302 cm$^{-1}$.  This mode
represents the vibrations of the chain oxygen atoms along the {\bf
a}-axis and appears in CuO at the same frequency \cite {a24}. The
mode at 548 cm$^{-1}$ is breathing mode of O$_1$ oxygen ladder
atoms.  Similar mode appears in SrCuO$_2$ at 543 cm$^{-1}$.  The
second oxygen ladder atom
 (O$_2$), Fig. 1,
is situated in the center of inversion ($D_{2h}$ site symmetry)
and has no Raman activity.  The highest frequency Raman mode in
Figs. 5(a)-(b), at about 582 cm$^{-1}$, is caused by the chain
oxygen vibrations along the {\bf c}-axis.
 Corresponding mode
appears in CuO at 633 cm$^{-1}$ \cite {a24} and in CuGeO$_3$ at
594 cm$^{-1}$ \cite {a25}. The vibrations of Sr atoms, with
frequency of 188 cm$^{-1}$ as in SrCuO$_2$, are not observed in
the spectra.  Finally, the A$_g$ mode at 153 cm$^{-1}$, see
 Fig. 8, originates from the vibrations of the Cu atom in chains. Again, similar
  mode is
found in YBa$_2$Cu$_4$O$_8$ at 153 cm$^{-1}$.

 By lowering
temperature Raman peaks narrow and at about T=150 K the new modes
appear (Fig. 5 and Table II). Similar effects are found in the IR
spectra as well. This temperature coincides with the charge
ordering temperature established in the NMR and neutron scattering
experiments \cite {a7,a9}. NMR study \cite {a7} showed the
splitting of signal from Cu$^{3+}$ ions into two peaks below 200K
suggesting the occurrence of the charge ordering. This effect is
confirmed by Cox {\it et al.} \cite {a11}.  They measured
synchrotron x-ray scattering on Sr$_{14}$Cu$_{24}$O$_{41}$ single
crystals and showed the appearance of the satellite peaks at (00l)
positions. The results
 are interpreted in terms of a charge-ordered model involving both dimerization
between two-nearest-neighbors of Cu$^{2+}$ ions surrounding a
Cu$^{3+}$ ion on a Zhang-Rice singlet site, and dimerization
between nearest-neighbors of Cu$^{2+}$ ions.

Fig. 8(a) shows the (cc) polarized low temperature (T=10 K) Raman
spectra of Sr$_{14}$Cu$_{24}$O$_{41}$ in the 125 - 750 cm$^{-1}$
spectral region, excited with 647.1 nm (1.91 eV) and 488 nm (2.54
eV) energies. There are a remarkable difference between Raman
spectra for these excitation lines which appears due to resonance
effects. Namely, both lines are very close to gap energies for
polarization along c-axis, see Fig. 3. The reflectivity spectra
measured at 10 K for the {\bf E}$||${\bf c} and the {\bf
E}$||${\bf a} polarizations are given in Figs. 8(b) and 8(c),
respectively. In order to compare Raman with IR data we shown in
the same figure the $\epsilon_2(\omega)$ and the
$-Im[1/\epsilon(\omega)]$ spectra. These spectra are obtained
using Kramers-Kronig analysis of reflectivity data. The TO and LO
mode frequencies are obtained as peak positions of the
$\epsilon_2(\omega)$ and the $-Im[1/\epsilon(\omega)]$,
respectively. For most of all Raman active modes we found theirs
infrared counterparts (some of them are denoted by vertical lines)
either for the {\bf E}$||${\bf a} or the {\bf E}$||${\bf c}
polarizations. The appearance of similar lines in IR and Raman
spectra, if not being the consequence of the symmetry, may also be
attributed to the resonant conditions. It is well documented \cite
{a26,a27} that Raman forbidden IR active LO modes appear in the
Raman spectra of the insulating Cu-O based materials for the laser
line energies close to gap values.  The appearance of these modes
in the Raman spectra is explained by Fr\"ohlich interaction \cite
{a26}. Here, since the 647.1 nm line is very close to the gap
values (1.86 eV, see Fig. 3) one can expected that the IR LO modes
appear in the Raman spectra of Sr$_{14}$Cu$_{24}$O$_{41}$ as well.
Such effect is usually accompanied by the observation of the
strong phonon overtones, as we shown in Fig. 6(a).
 All modes  with energies higher then 700
cm$^{-1}$ are in fact the second order combinations (overtones) of
the low-energy modes. The assignment and the frequencies of these
modes are given in Table III. Therefore, the properties of the
modes observed in our spectra may be understood in terms of the
available symmetry combined with resonant effects, thus making
proper identification of the low-temperature phonons practically
impossible without detailed structural analysis.

Now we focus on magnetic properties. As it is mentioned earlier,
the neutron scattering and NMR measurements estimated the
spin-ladder gap value to be at $\Delta_L$ = 32.5 meV  (268
cm$^{-1}$) \cite {a8} or at 40.5 meV (326 cm$^{-1}$) \cite {a28},
respectively. Because of that, we paid special attention to the
200-350 cm$^{-1}$ spectral range, Fig. 5 (left panel).  By
lowering temperature we observe the appearance of the new modes at
262, 293, and 317 cm$^{-1}$. At the same time we find the modes
with the same energy in the low-temperature IR spectra, Figs.
8(b)-(c). The 262 cm$^{-1}$ mode is very close in energy to the
magnetic gap, thus possibly one-magnon excitation, as proposed by
Sugai {\it et al.}\cite {a15}. However, the origin of the
one-magnon excitation in the light scattering process usually
comes from the spin-orbit interaction, which is found to be very
small in transition metal oxides due to quenched orbital momentum
of the transition metal ions. Moreover, below 100 K, this mode has
nearly the same temperature dependency of the frequency and
intensity, like all other low-temperature modes and we identify
them as zone edge phonons, which become Raman active due to the
zone folding effect caused by the charge ordering transition \cite
{a7}. Yet another type of magnetic excitations is expected to
appear in the Raman spectra of two-leg-ladders at energies close
to $2  \Delta_L \sim$ 534 cm$^{-1}$ \cite {a29}. The mode at 498
cm$^{-1}$ (see Fig. 5 (g), right panel) shows a typical asymmetric
shape with a tail towards high-frequencies, as expected for the
onset of the two-magnon continuum. Its energy difference from
$2\Delta_L$ could be the magnon binding energy. However, as it is
shown in Fig. 8(b), this mode is positioned between TO and LO
frequencies of
 very
strong IR active mode in the {\bf E}$||${\bf
c} spectra. This mode also shows a strong resonant enhancement,
 see Fig. 8 (a). Therefore, at this
stage, it is hard to make definite conclusions about the origin of
this mode and further experiments are needed to clarify this
issue. Similar discussion holds for the spectral range around the
twice the spin-gap value associated with a chains, $2\Delta_C \sim
$180 cm$^{-1}$ \cite{a8}, where continuum-like feature is also
found in the Raman spectra.

 Finally, we discuss the modes in the spectral range above 1500 cm$^{-1}$.
The strongest mode in the spectra is centered at about 2840
cm$^{-1}$ for the 488 nm excitation and at about 2880 cm$^{-1}$
for the 514.5 nm excitation line. This feature decrease in
intensity and shifts to lower energies at higher temperatures,
Fig. 7. The same structure is already observed in many copper
oxides at similar frequencies \cite{a27,a29}. Thus, all observed
effects indicate the two-magnon origin of this mode. The energy of
the two-magnon mode, associated with a top of the magnon brunch,
in copper oxide insulators is about 3J, where J represents the
exchange interaction.

In the case of the two-leg-ladders, its energy position for
different polarized configurations in the Raman spectra may be
used to estimate the exchange parameters parallel ($J$) and
perpendicular ($J_0$) to the ladders \cite{a30}. Since the energy
position of the two-magnon peak is the same for (cc) and (aa)
polarizations, Fig. 6(c), such an analysis suggests that $J=J_0$.
This conclusion is fully in agreement with Raman scattering data
of Ref.\cite{a15} and recent high-energy neutron scattering
measurements\cite{a31}, but inconsistent with previous neutron
\cite{a8}, NMR\cite{a6} and magnetization measurements, which
estimated $J_0/J$ ratio to be between 0.5 and 0.8. The discrepancy
may be related to the fact that in the previous measurements, the
high-energy magnetic excitations were not observed directly, but
estimated indirectly from the low-energy spin gap measurements
assuming an ideal model \cite{a32}. Thus, the Raman scattering is
more direct method to obtain $J_0/J$ ratio. The two-magnon mode at
2880 cm$^{-1}$ gives exchange energy $J$=119 meV, very close to
the neutron scattering value $J$=130 meV \cite {a8}.

It is also interesting to note that the two-magnon mode is
asymmetric with the spectral weight shifted to higher frequencies.
Such a spectral shape of the two-magnon mode can be a consequence
of the resonance \cite {a30}, or it can be related to the
bound-hole-pair effects \cite {a15}. Still, further experiments on
the hole doped crystals are necessary to clarify this point.

Let us consider Fig. 6(b), where a weak structure appears in (cc)
spectra at about 1920 cm$^{-1}$.  Its energy is exactly equal to
2J and varies with a laser line frequency in a similar way as the
two-magnon mode at 2880 cm$^{-1}$. The energy shift of this mode
as a function of temperature was not seen because of its very low
intensity, thus leaving the origin of this mode as an open
question. In addition to the 1920 cm$^{-1}$ mode, we found a weak
structure at about 1700 cm$^{-1}$, as well. This mode does not
possess any noticeable temperature dependencies of energy and
intensity. We concluded that this mode is an overtone phonon mode.
Its frequency can be represented as the third order of the 568
cm$^{-1}$ mode ($3c_3$, see Table II).

Finally, we discuss the electron energy dispersions and density of
states, which are calculated using the model described in Sect.
III. First, we analyze the influence of the energy transfer
(hopping) to the correlation gap. Fig. 9 shows the correlation gap
$\Delta_{corr}$ vs. the Anderson-Hubbard parameter $U$. The
plotted curves are calculated using Eq.(5) and $J=4t^2/U$=0.13 eV
for different hopping energy ratios $t_{xy}/t$ and $t_0/t$. From
Fig. 9 we conclude that the main influence to the correlation gap
value comes from the Anderson-Hubbard parameter U. Because of
that, by knowing the electronic gap from the ellipsometric or
optical absorption measurements and the exchange energy J from the
Raman spectroscopy we can determine the hopping parameters and the
onsite electron-electron repulsion $U$.
 However, relative
ratio of the hopping energies perpendicular and parallel to the
legs as well as the interladder hopping, does not influence much
the correlation gap . Namely, a decrease of the transfer energy
along the rungs from $t_0/t$=1 to $t_0/t$=0.5 increases
$\Delta_{corr}$ for about 6\%. Also, an increase of the
interladder hopping $t_{xy}$ from 0 to 10\% of the hoping value
$t$ along the legs, produces a decrease of $\Delta_{corr}$ for
about 2.5\%. Therefore, the interladder effects on electronic
structure of Sr$_{14}$Cu$_{24}$O$_{41}$ are found to be negligibly
small even though the distance between the neighboring ladders is
short (see Fig. 1). The correlation gap is observed at 1.4 eV.
This value is determined as a maximum of dielectric function
$\epsilon_2(\omega)$ obtained from KKA of the unpolarized
reflectivity data, see Inset (b) of Fig. 3(b). Using this value
and the fact that $t_0/t=1$ (comes form $J_0/J=1$), $t_{xy}=0$ we
obtained U=2.1 eV. Similar values for the correlation gap and U
has been also found in $SrCuO_2$ \cite {a34}.

Energy dispersion, shown in Inset of Fig. 2, allows us to assign
the 1.86 eV peak in Fig. 3(a) to the gap value ($\Delta_1$=1.87
eV) between bonding and antibonding bands at the Z-point.  Also,
2.4 eV peak from the ellipsometric measurements corresponds to the
gap from the lowest occupied band ($L_6$, Fig. 2) to the highest
empty band at Z-point of Brillouine zone ($\Delta_2$=2.4 eV). By
comparison of our measured and calculated gap values, with
previously published results \cite {a17,a35}, we found that our
$t_0/t$ ratio is close to the ratio determined in \cite {a17}.
Mizuno {\it et al.}, \cite {a35}, calculated the optical
conductivity for small clusters, simulating the ladders and the
chains.  They obtained the gap for the ladder at about 1.7 eV
while the contribution from the chains mainly emerges at a higher
energy showing the large spectral weight at around 2.6 eV. These
values are very close to our experimental results. Therefore, the
peaks at 1.86, 2.4 and 2.5 eV, see Fig. 3(b), may correspond to
the ladders and the chains, respectively.

In conclusion, we studied the optical properties of the
Sr$_{14}$Cu$_{24}$O$_{41}$ single crystal. The lattice vibrations
are analyzed using the far-infrared reflectivity and Raman
scattering measurements in the wide frequency and temperature
range. At temperatures below 150 K the new IR and Raman modes
appear, presumably due to the charge ordering. The two-magnon
excitations are found in the Raman spectra which could be related
to minimal ($2 \Delta$) and maximal (twice the top of the magnon
brunch) magnon energy. The exchange constants along the legs and
rungs of the ladders are found to be the same, $J=J_0 \sim$ 120
meV. The optical reflectivity and the ellipsometric measurements
are used to study the charge dynamics. The gap values of 1.4 eV,
1.86 eV (2.5 eV) for the ladders (chains) along the {\bf c}-axis
and 2.4 eV along the {\bf a}-axis are found. These results are
analyzed using tight-binding approach for the correlated
electrons. The correlation gap value of 1.4 eV is calculated with
the transfer energy (hopping)
 parameters $t$=t$_{0}$=0.26 eV, along and perpendicular to the legs,
  and U=2.1 eV, as a Coulomb repulsion.

\section{Acknowledgment}

We thank W. K\"onig for low-temperature infrared measurements.
Z.V.P., V.A.I and O.P.K acknowledge support from the Research
Council of the K.U.  Leuven and DWTC.  The work at the K.U. Leuven
is supported by the Belgian IUAP and Flemish FWO and GOA Programs.
M.J.K thanks Roman Herzog - AvH for partial financial support.

\clearpage

\begin{figure}
\caption {Schematic representation of the Sr$_{14}$Cu$_{24}$O$_{41}$
 crystal structure in
the (a) (010) and (b) (001) plane. }
\label{fig1}
\end{figure}

\begin{figure}
\caption {The electron density of states as a function of
dimensionless
 energies $\xi/2t$. Inset: The tight-binding dispersions for correlated
 electrons in Sr$_{14}$Cu$_{24}$O$_{41}$ with parameters $t$=$t_0$=0.26 eV, $t_{xy}$=0.026 eV.
 The momenta are given in units
 $|p_x\sqrt{2}|$=$|p_z|$=$\pi$ of the Brillouine zone boundaries, the Fermi energy
 $E_F=0$ is inside of the correlation gap.}
\label{fig2}
\end{figure}

\begin{figure}
\caption {Room temperature imaginary part ($\epsilon_2$) of the
pseudodielectric function of Sr$_{14}$Cu$_{24}$O$_{41}$. The
spectra of the (010) surface taken with a) a-axis, b) c-axis,
parallel to the plane of incidence. Inset: (a) Reflectivity
spectra obtained by calculation using $\epsilon_1$ and
$\epsilon_2$; (b) unpolarized reflectivity spectrum measured at
room temperature. } \label{fig3}
\end{figure}

\begin{figure}
\caption {Room temperature polarized far-infrared reflectivity
spectra of Sr$_{14}$Cu$_{24}$O$_{41}$ single crystal for (a) the
{\bf E}$||${\bf a} and (b) the {\bf E}$||${\bf c} polarizations.
The experimental values are given by the open circles. The solid
lines represent the calculated spectra obtained by fitting
procedure described in the text. } \label{fig4}
\end{figure}
\begin{figure}
\caption {Temperature dependent Raman scattering spectra for (aa)
polarized (a) and (cc) polarized (b)-(g) configurations.
$\lambda_L$ = 514.5 nm.  } \label{fig5}
\end{figure}
\begin{figure}
\caption {Raman spectra measured at T=8 K in the 700-3300
cm$^{-1}$ spectral range. } \label{fig6}
\end{figure}
\begin{figure}
\caption {Frequency and intensity dependencies of the two-magnon
 mode at about 2840 cm$^{-1}$ as a function of
 temperature. }\label{fig7}

\end{figure}
\begin{figure}
\caption {(a) The (cc) polarized Raman spectra measured at T=10 K
with 647.1 and 488 nm excitation lines. (b) Polarized far-infrared
reflectivity spectra measured at T=10 K for $E||c$ (b) and $E||a$
(c) polarizations. Dielectric functions $\epsilon_2(\omega)$ and
$-Im[1/\epsilon(\omega)]$ are obtained using KKA of reflectivity
data.} \label{fig8}
\end{figure}
\begin{figure}
\caption {The correlation gap vs. U parameter dependence for
different transfer energy ratios. } \label{fig9}
\end{figure}

\begin{table}
\caption{Oscillator fit parameters (in cm$^{-1}$) of the
reflectivity data, T=300 K.}
\begin{tabular}{cccccccc}
\tableline

Polarization & $\omega_{TO}$ & $\gamma_{TO}$ & $\omega_{LO}$ &
$\gamma_{LO}$ & $\omega_p$ & $\tau^{-1}$ & $\varepsilon_\infty$
\\ \tableline & 164 & 6 & 164.6  & 5.5 & \\ & 194 & 6 & 210 & 14 &
\\ & 219 & 10 & 223 & 10 & \\ & 249 & 14 & 250 & 14 & \\ ${\rm\bf
E\parallel a}$ & 283.5 & 14 & 284 & 14 & 1000 & 4000 & 4.8
\\ & 310.4 & 8 & 311 & 8 & \\ & 554 & 16 & 585 & 28 & \\ & 623 &
19 & 676 & 22 & \\ \tableline & 135 & 6 & 140  & 10 & \\ & 148 & 5
& 150 & 5 & \\ & 253 & 12 & 257 & 17 & \\ & 293 & 18 & 300 & 14 &
\\ ${\rm\bf E\parallel c}$ & 345 & 26 & 364 & 10 & 4000 & 18200
& 4 \\ & 486 & 20 & 502 & 20 & \\ & 540 & 25 & 568 & 17 & \\ & 596
& 20 & 600 & 17 & \\ & 620 & 30 & 629 & 14 & \\ \tableline
\end{tabular}
\end{table}

\begin{table}
\caption{Frequencies (in cm$^{-1}$) of Raman active modes measured
at T=8 K with $\lambda_L$= 488 nm, 514.5 nm and 647.1 nm, (Figs.
5(g) and 8(a)). By * we denoted the modes influenced by charge
ordering. }

\begin{tabular}{cccc}
\tableline 488 nm & 514.5 nm & 647.1 nm & Remark \\ \tableline - &
- & 115 &
-
\\ 154& - & 153 & $a_1$ \\ 197 & - & - & -\\ 208 & 213* & 208 & $a_2$ \\
224& 224& 223&-\\ -&-&230&-\\ -&237*&238&-\\ 251&254&-&-\\
-&265*&-&-\\ -&268*&-&-\\ -&293*&-&-\\ 304&304&-&-\\ - &316& 316 &
$a_3$
\\ -&- & 359 & $c_1$
\\ -&-&381&-\\
500&497*& 496 & $c_2$
\\ -&516*&-&-\\
532&535*&-&-\\ 553&556&-&-\\ - &568*& 568 & $c_3$ \\ 586&585 & 585
& $a_4$
\\ 632 & 630 &-&-
\\-&-&652&$a_5$\\  \tableline

\end{tabular}
\end{table}

\begin{table}
\caption{Mode frequencies (in cm$^{-1}$) observed in Raman spectra
measured with $\lambda_L$=647.1 nm, Fig.6(a). }

\begin{tabular}{ccc}
\tableline No. of peaks & Frequency & Remark \\ \tableline  1 &
735 & $a_1+a_4$ \\ 2 & 758 & -
\\ 3 & 805 & $a_1+a_5$ \\ 4 & 860 & $a_2+a_5$ \\ 5 & 946 &
$c_1+a_4$ \\ 6 & 1005 & $c_1+a_5$ \\ 7 & 1069 & $c_2+c_3$ \\ 8 &
1152 & $c_2+a_5$ \\ 9 & 1164 & $2a_4$ \\ 10 & 1132 & $a_4+a_5$
\\ 11 & 1300 & $2a_5$ \\ \tableline

\end{tabular}
\end{table}

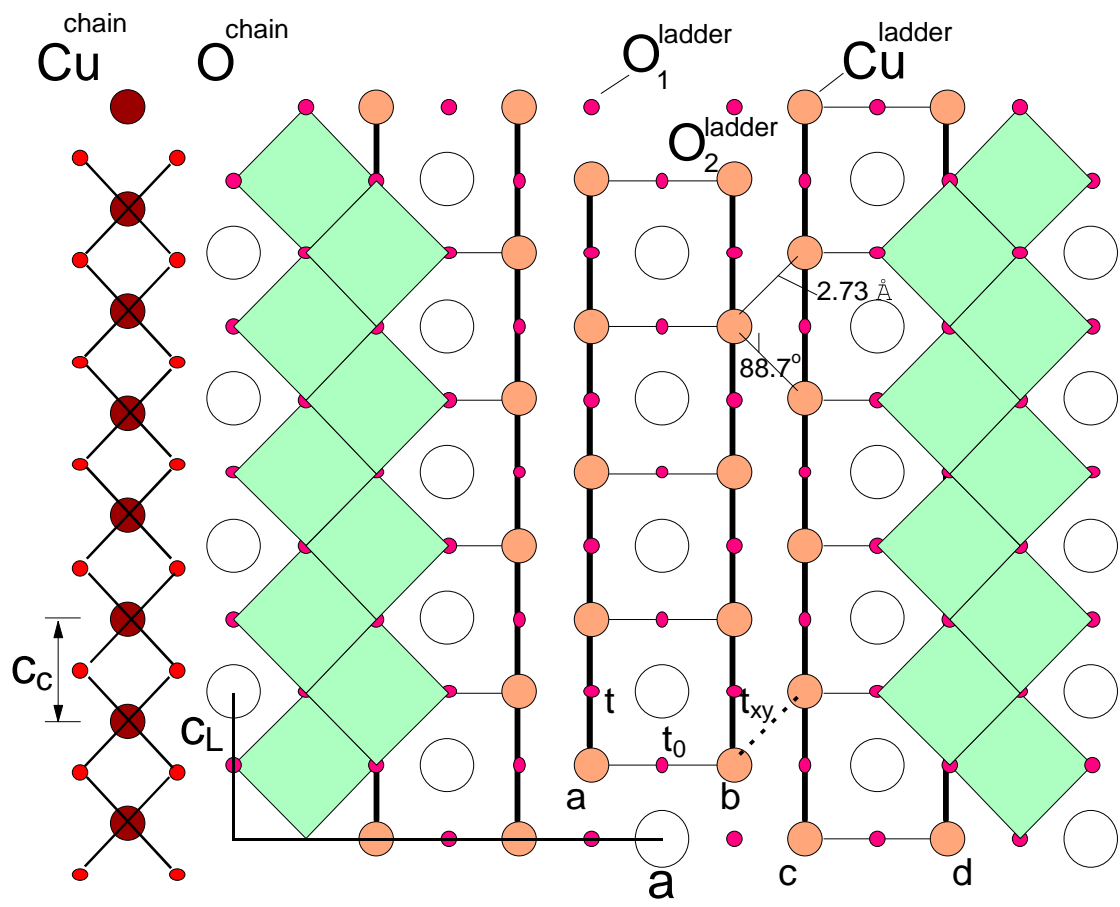
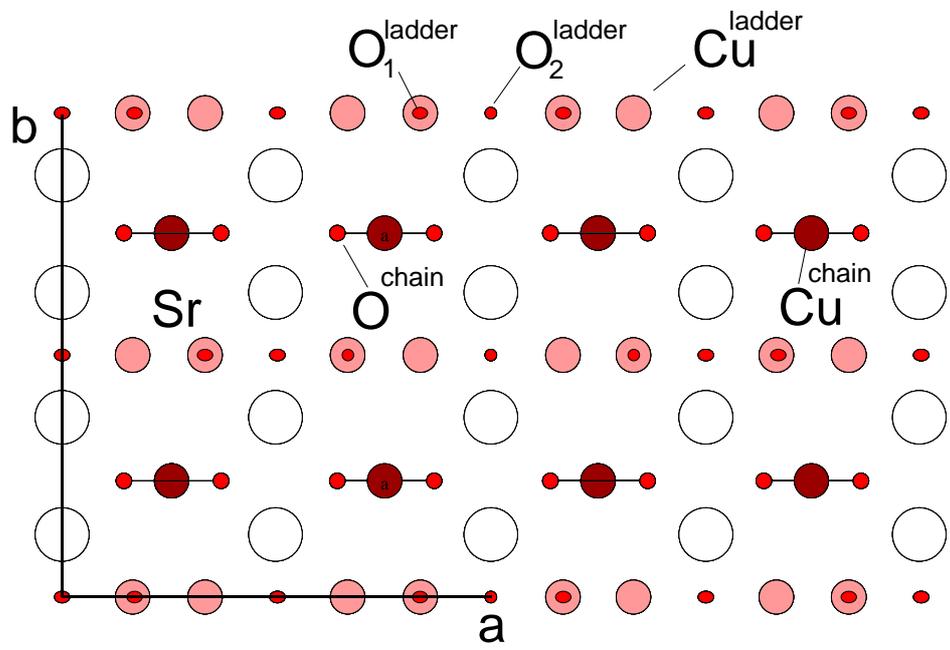
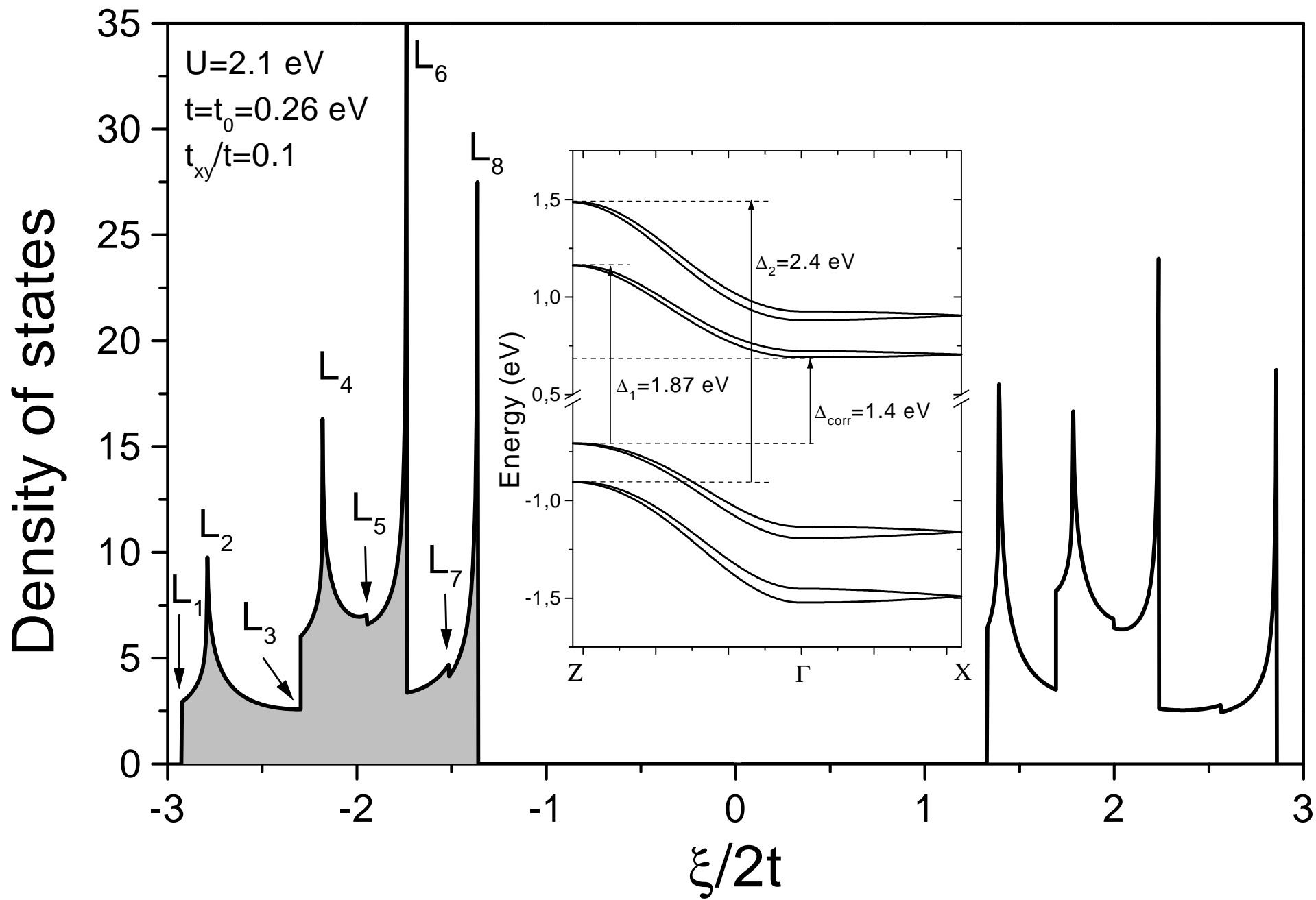
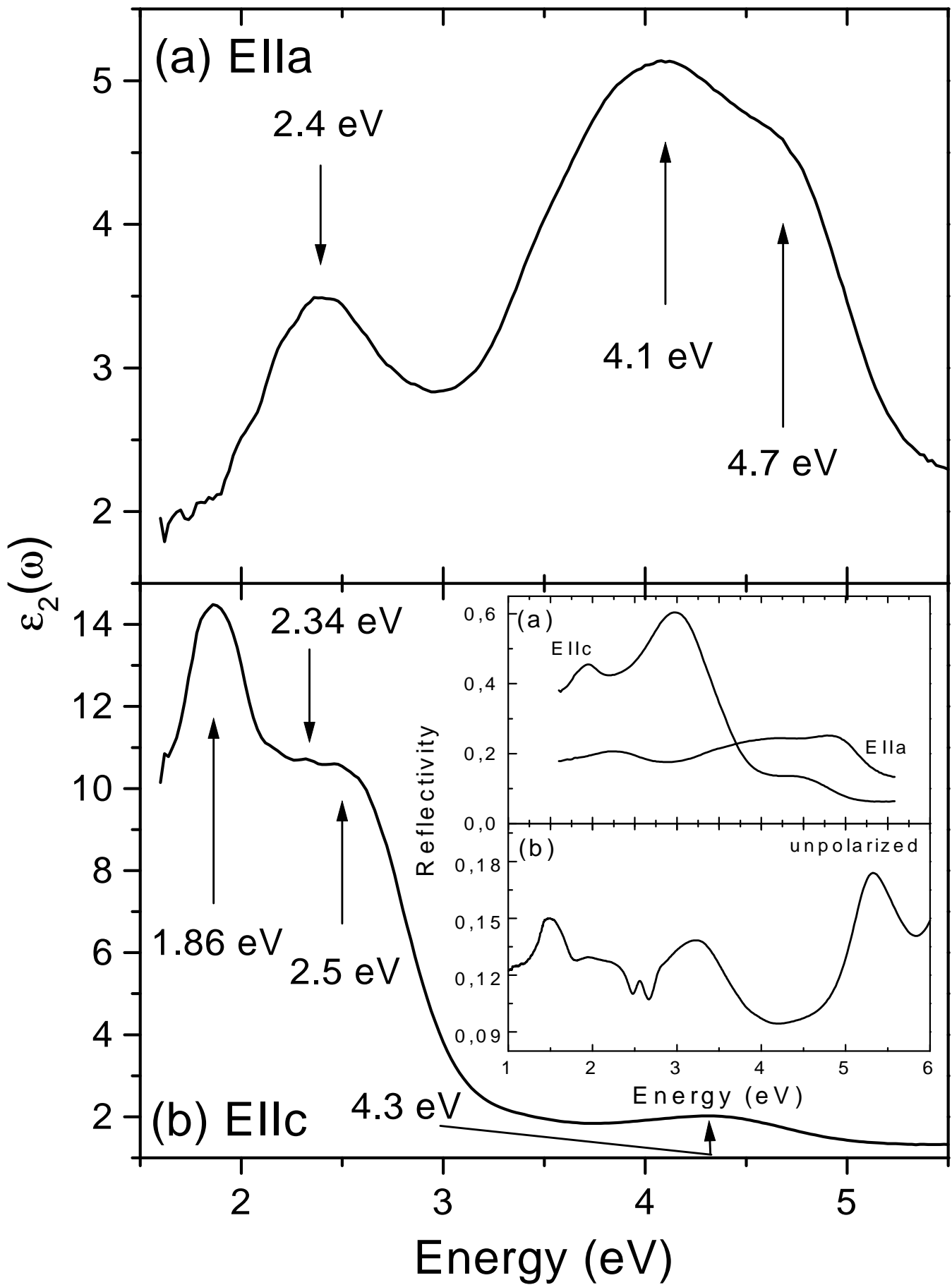
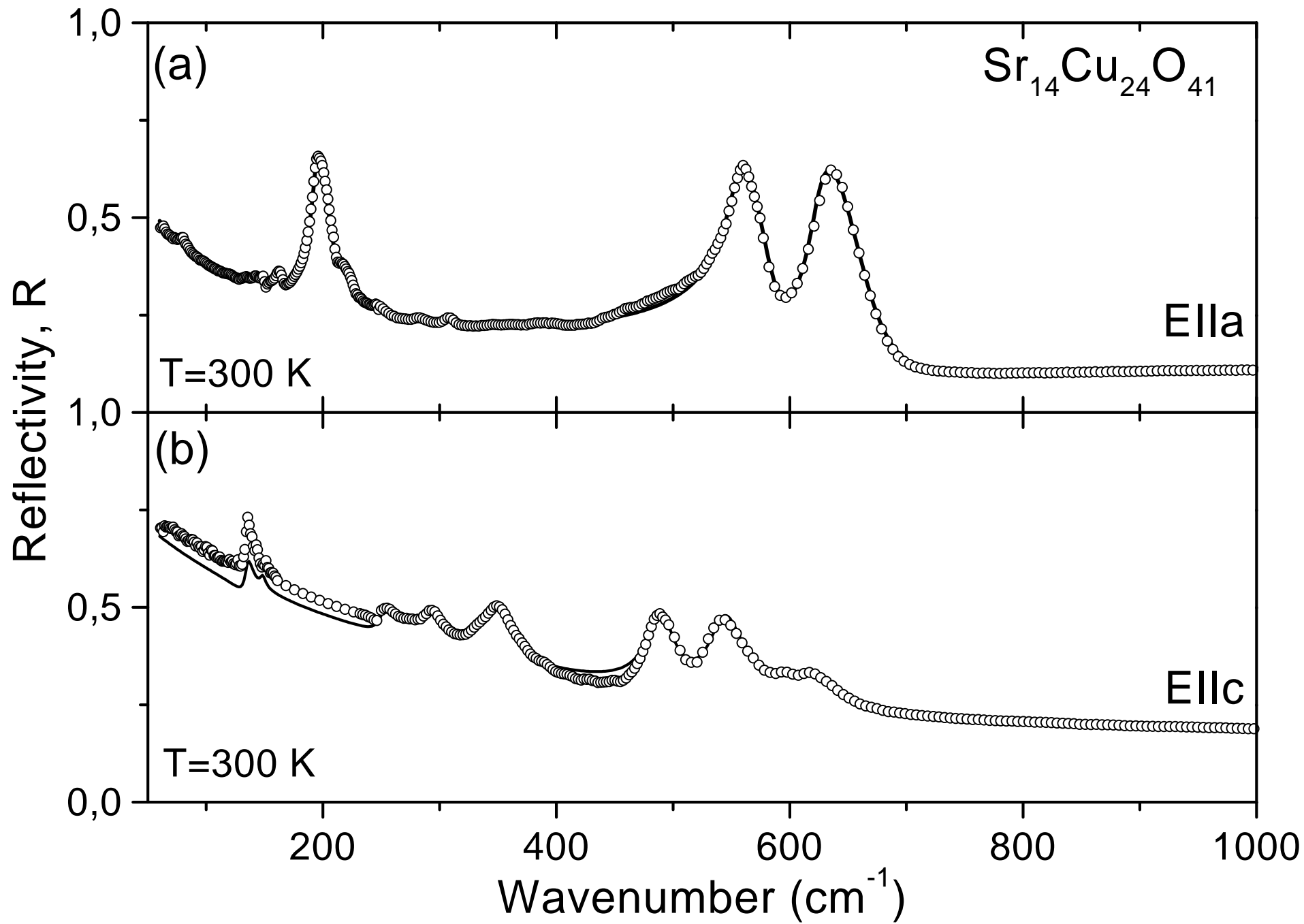
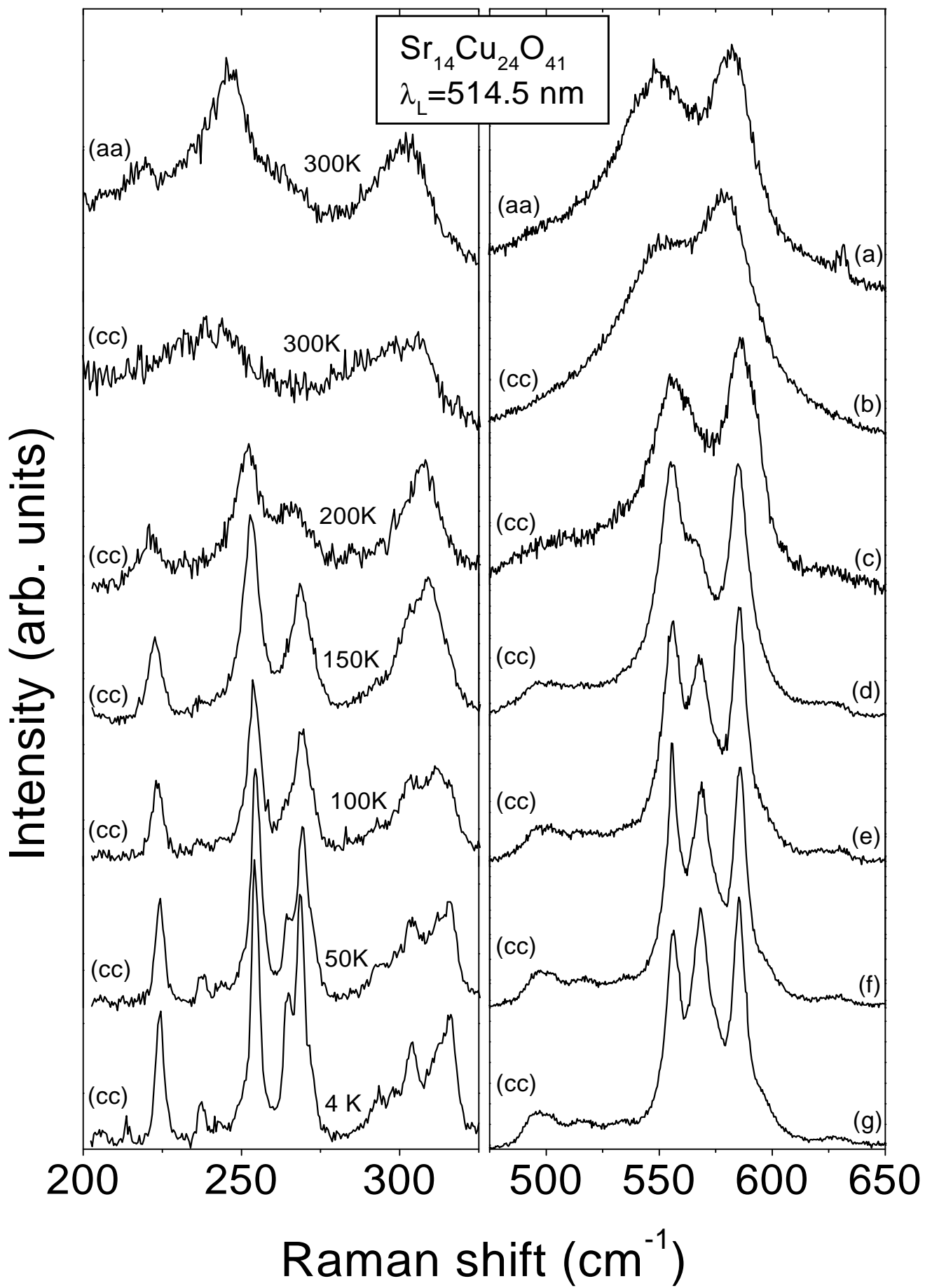
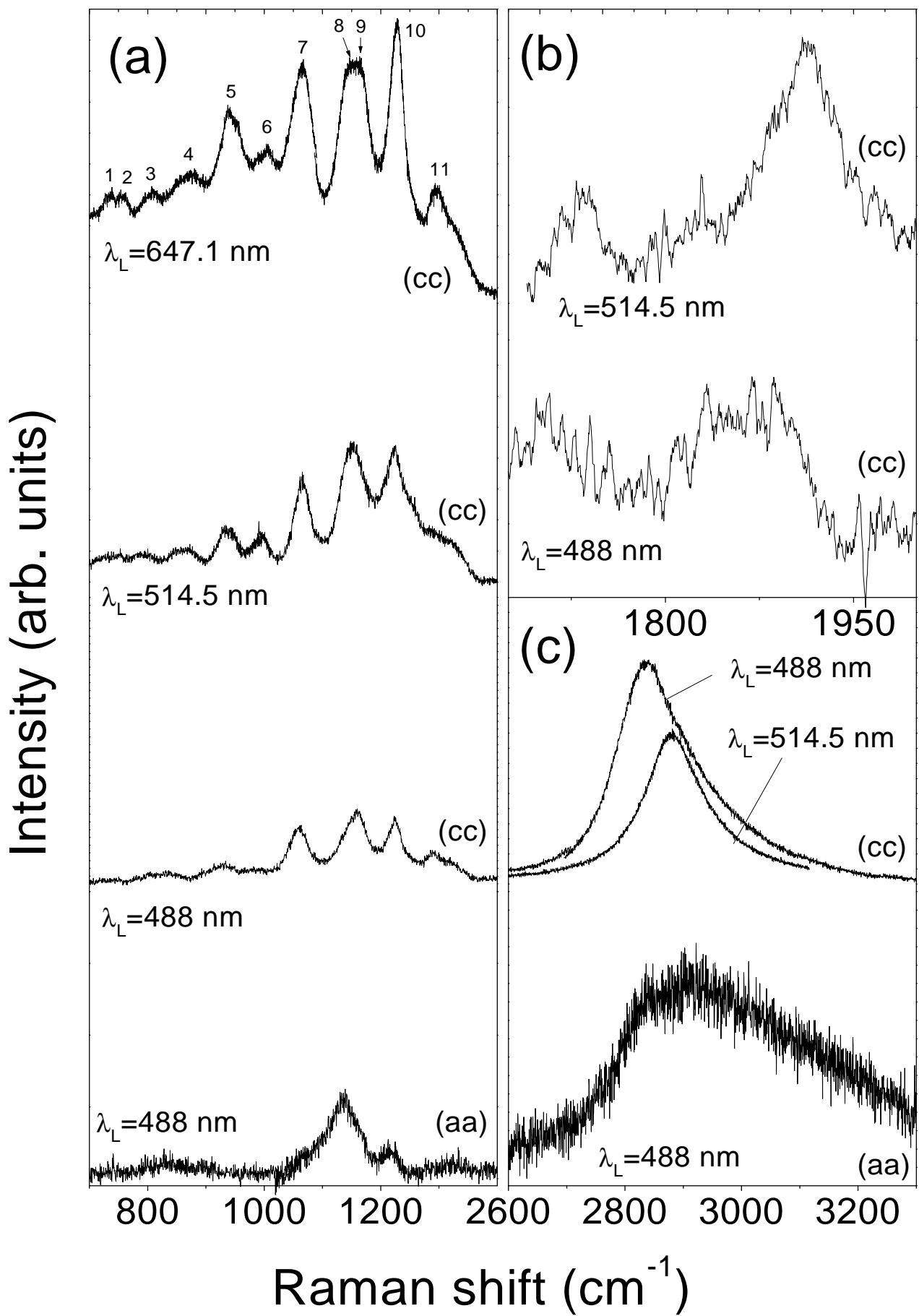
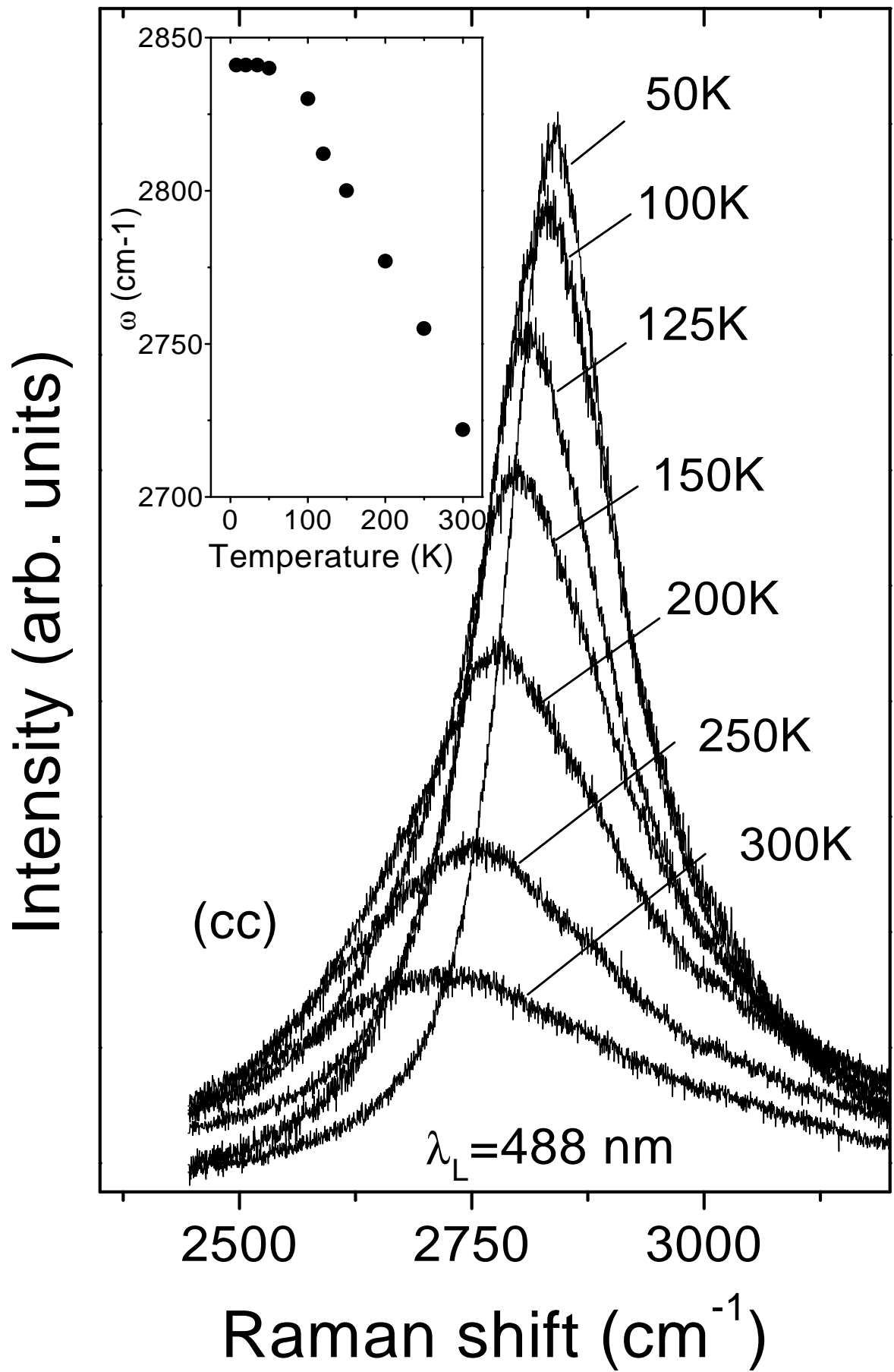
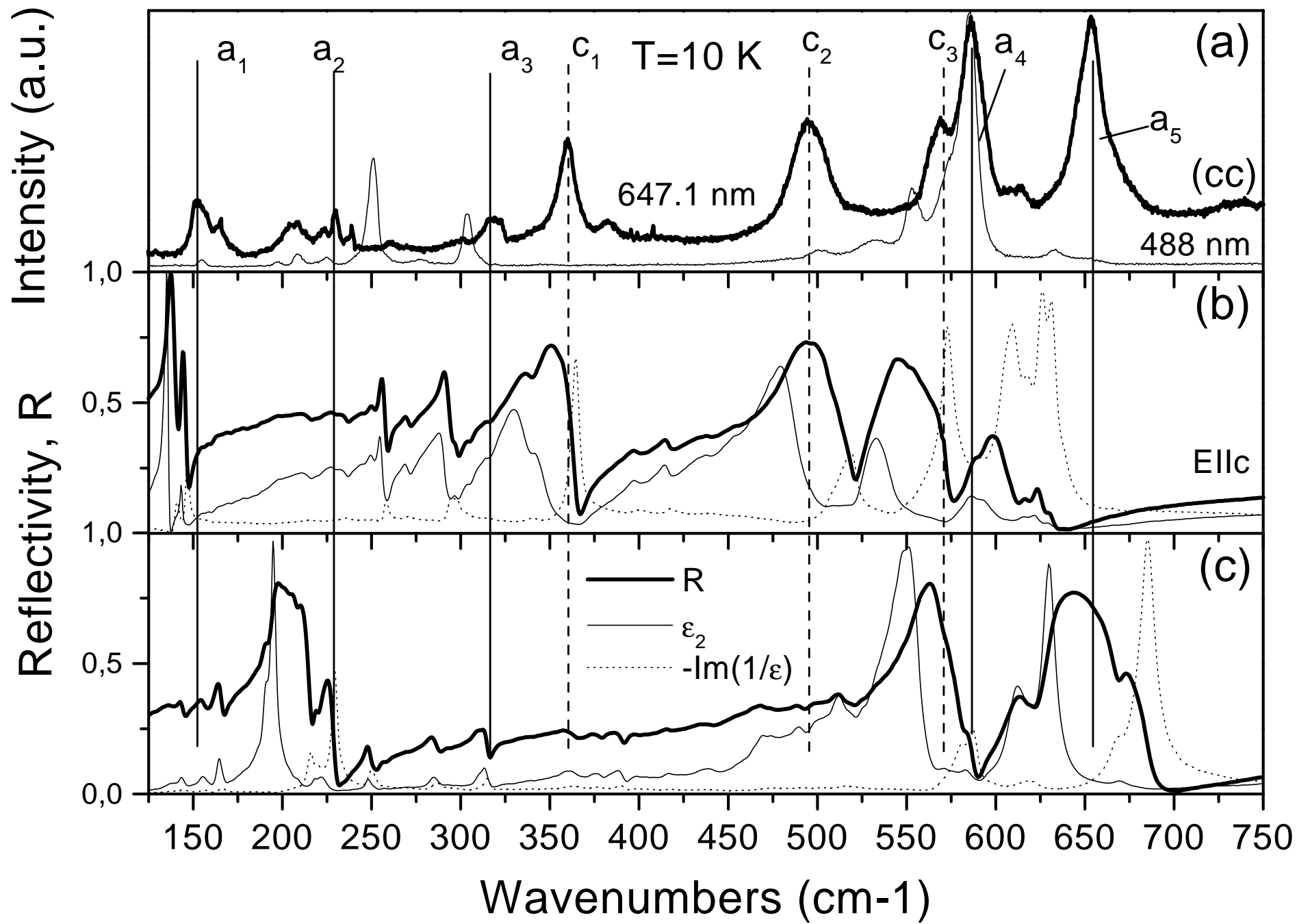
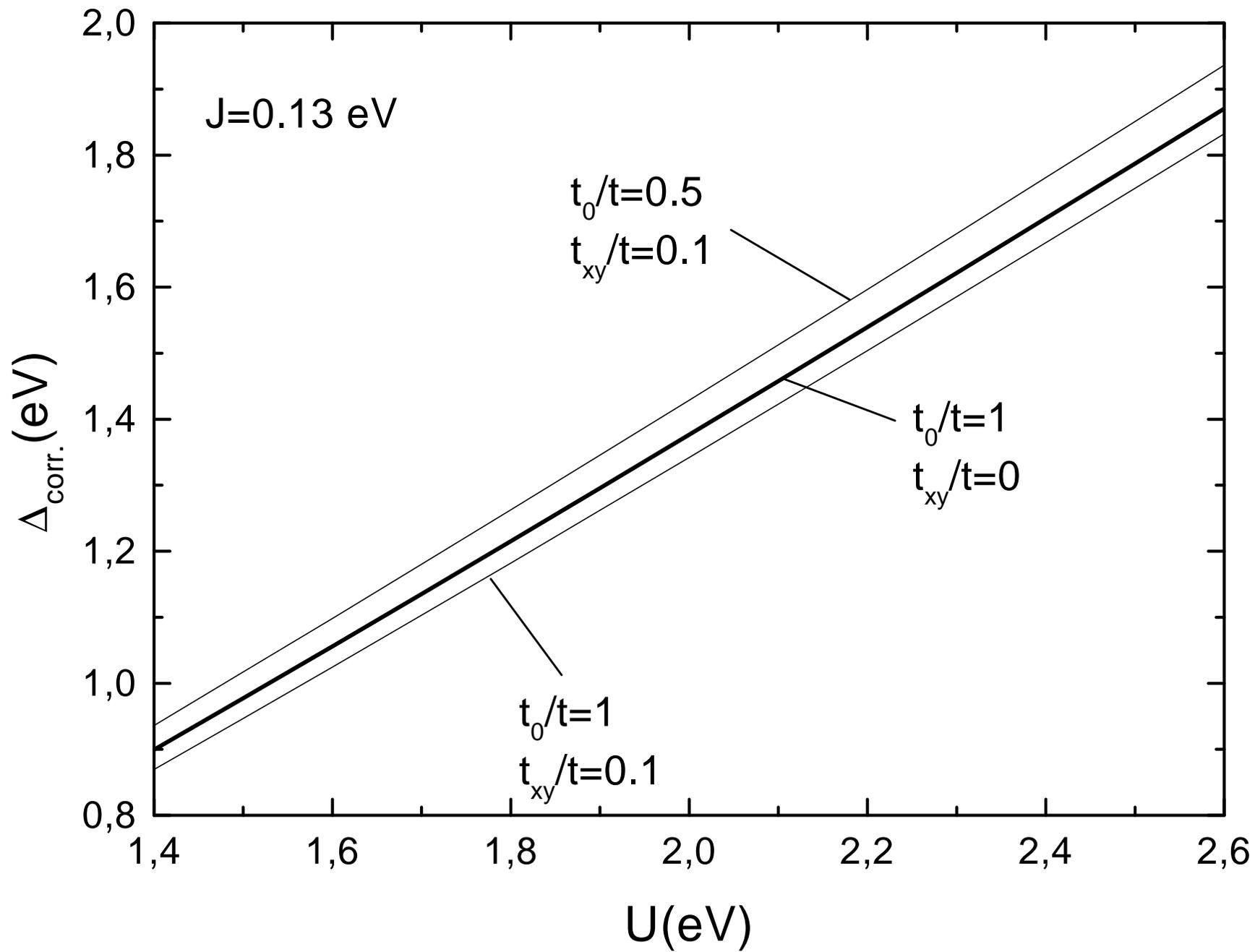
\end{document}